\documentclass{article}
\usepackage{spconf,amsmath,epsfig,graphicx,subfig,tabularx,booktabs,threeparttable,float,cite,hyperref}

\let\OLDthebibliography\thebibliography
\renewcommand\thebibliography[1]{
  \OLDthebibliography{#1}
  \setlength{\parskip}{0pt}
  \setlength{\itemsep}{0pt plus 0.3ex}
}

\begin{document}
\topmargin=0mm
\sloppy
\def\x{{\mathbf x}}
\def\L{{\cal L}}

\title{UNSUPERVISED QUANTIZED PROSODY REPRESENTATION FOR CONTROLLABLE SPEECH SYNTHESIS}
%
\name{Yutian Wang$^{1}$, Yuankun Xie$^{1}$, Kun Zhao, Hui Wang$^{1\ast}$ \thanks{*Corresponding author}, Qin Zhang$^{1}$ \thanks{This work was supported in part by the National Natural	Science Foundation of China under Grant 61631016}}
\address{1.State Key Laboratory of Media Convergence and Communication, Communication University of China,\\ Beijing, China}

\maketitle

\begin{abstract}
In this paper, we propose a novel prosody disentangle method for prosodic Text-to-Speech (TTS) model, which introduces the vector quantization (VQ) method to the auxiliary prosody encoder to obtain the decomposed prosody representations in an unsupervised manner. Rely on its advantages, the speaking styles, such as pitch, speaking velocity, local pitch variance, etc., are decomposed automatically into the latent quantize vectors. We also investigate the internal mechanism of VQ disentangle process by means of a latent variables counter and find that higher value dimensions usually represent prosody information. Experiments show that our model can control the speaking styles of synthesis results by directly manipulating the latent variables. The objective and subjective evaluations illustrated that our model outperforms the popular models.
\end{abstract}
\section{Introduction}
\label{sec:intro}

With the development of deep learning in recent years, TTS model has also experienced rapid growth. The neural TTS models, such as Deep Voice\cite{Arik2017}, Tacotron\cite{Wang2017,Shen2018}, Fastspeech\cite{Ren2019,Ren2021}, etc. have greatly improved the naturalness and computational efficiency of speech synthesis.
However, most of the TTS models can only learn the fixed prosody from the training dataset, and can not change the prosody by adjusting the model parameters. To overcome this shortcoming, several improved models are proposed  \cite{Wang2018,Kwon2019,Wu2020,Akuzawa2018,Sun2020,Zhang2019}. Most of them use an auxiliary encoder to encode the reference speech into prosody embedding vector in unsupervised way, and generate similar prosodic speech by conventional TTS model under the condition of  prosody embedding vector. Therefore, the core problem becomes how to extract prosodic information from reference speech while avoiding the interference of other information like content, speaker ID, etc. \\
The GST-Tacotron model \cite{Wang2018,Kwon2019,Wu2020}  use a set of randomly initialized tokens called Global Style Tokens (GSTs) to represent the prosody and synthesize speech with Tacotron model. GSTs are updated as the query vector of an attention module which can disentangle the reference speech characters into these tokens. The prosody of synthesized speech can be modified by adjusting the value of GSTs. However, the disentanglement performance of GST is not ideal enough, so that the generated results is not satisfactory. In addition, the relationship between GSTs and prosody characters is not clear, which needs to be tested manually. If there are too many GSTs, it is a difficult task to complete.\\
To improve the performance of prosodic speech synthesize, some more complicated auxiliary encoders were proposed.  Inspired by GSTs, Hierarchical GST is presented to learn multiple-level disentangled representations \cite{An2019}. With the number of GST layers increases, it tends to learn a coarse to fine style decomposition. On the other hand, due to the self-supervised learning ability, Variational Auto-Encoder (VAE) is used as auxiliary encoder to force the latent representation to Gaussian distribution \cite{Akuzawa2018,Sun2020,Zhang2019}. With this advantage, the speaking styles are disentangled on the different dimensions of the Gaussian distribution which could independently control different style attributes, such as pitch, local pitch variation, speaking velocity, etc. Furthermore, a hierarchical latent representation is implemented by two layers VAE with Gaussian Mixture prior \cite{Hsu2019}, which can learn labeled and unlabeled attributes and control more detailed speaking styles. However, although their disentangle ability have been greatly improved, they are representation ability is still need to be improved. In other words, which dimension corresponds to which prosodic feature is not clear. \\
From the above description, there are two problems with existing models. First is how to extract prosodic information from reference speech while avoiding the interference of other information like content, speaker ID, etc. Second is how to enhance the representation ability of models and establish the relationship between latent variables and real prosodies.\\
Vector-Quantized Variational Auto-Encoder (VQ-VAE) \cite{Oord2017,Razavi2019} uses a special codebook mapping method to make the latent representations discretized, which provides a different kind of latent spaces form. In voice conversion task, it shows good disentangle ability of separating speech content and timbre\cite{Wu2020a}. Inspired this, in this paper, we propose a VQ auxiliary encoder which use the Vector-Quantized latent variable as the representation of prosody. Benefit from the discrete latent space, our model can archive better disentangle performance and representation ability, which allow us to directly control the prosody of the synthesized speech manually. 
We also investigate the internal mechanism of VQ disentangle process and find that it can automatically separate long-term and short-term information into different dimensions. We use a numerical counter to count the updated values of latent codebook in different dimensions. It is found that higher values usually represent prosody information.

\section{MODEL}
\label{sec:model}

\subsection{Auxiliary encoder}
\begin{figure}[htb]
	\begin{minipage}[b]{1.0\linewidth}
		\centering
		\centerline{\includegraphics[width=8.5cm]{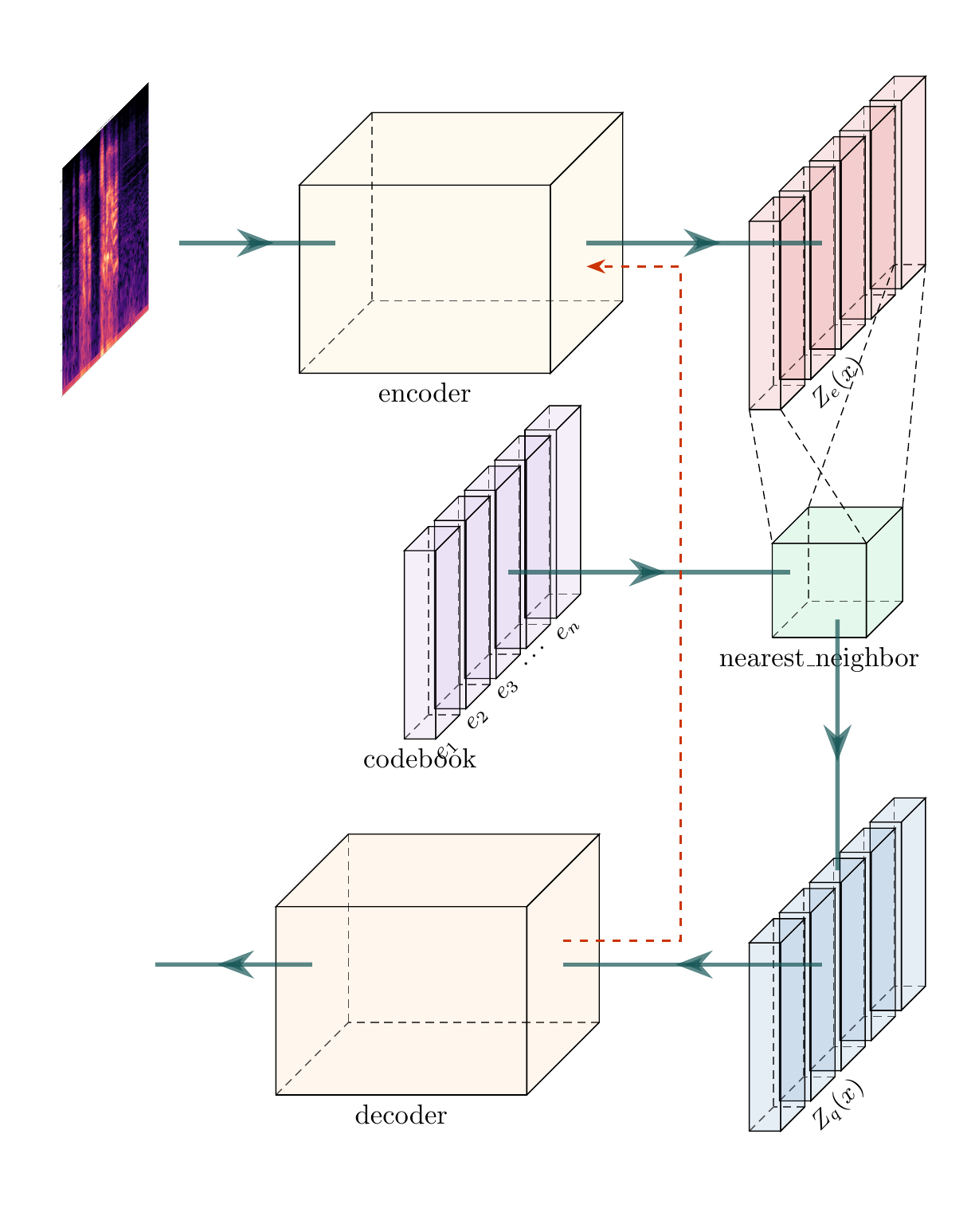}}
		\caption{The schematics of the vector quantization method. In the diagram, $Z_e (x)$ is a continuous variable and $e_j$ is random initialized discrete variables. The Nearest Neighbor method finds the corresponding vector $Z_q (x)$ and pass it to decoder. In the process of gradient updating, gradient information is directly transferred from decoder to encoder, which avoids the problem that the gradient cannot be calculated in VQ process, which shows as red dash line. }
		\label{vq_figure1}
	\end{minipage}
\end{figure}

In the auxiliary encoder, we use the vector quantization method to extract the prosody embedding from reference audio. At the beginning of the training process, the codebook is constructed by discrete embedding vectors $\{e_i  ,i=1,…,n\}$, which are randomly initialized. Then, the nearest neighbor lookup method is used to calculate the discrete latent variables  $Z_q (x)$ as given in Equation \ref{vq_basic_equation}, where $Z_e (x)$ is the output of the encoder and $Z_q (x)$   is the calculation result of the nearest neighbor method. According to Equation \ref{vq_basic_equation}, the quantized vector $Z_q (x)$ will be identical to the value of codebook component $e_j$, if the encoder output $Z_e (x)$ close to $e_j$. At last, the discrete latent representation $Z_q (x)$ will be passed to the decoder. The vector quantization process is shown in Figure \ref{vq_figure1}.  

\begin{equation}
	Z_q (x)=e_k, k={argmin}_j \|Z_e (x)-e_j \|_2   \label{vq_basic_equation}
\end{equation}

The loss function of the vector quantization process is defined as follows:
\begin{equation}
	\begin{aligned}
		L=&\log p(x|Z_q(x))+ \|{SG}[Z_e(x)]-e\|_2^2\\
		&+\beta\|Z_e(x)-{SG}[e]\|_2^{2}   \label{vq_loss}     
	\end{aligned}    
\end{equation}
The first term of Equation \ref{vq_loss} is a reconstruction loss between the input data and the reconstructed data. Besides that, the second and third terms are about vector quantization. The second term evaluates the distance between latent vectors and codebook vectors using $L_2$ Norm to make codebook vectors close to the encoder output $Z_e (x)$. $ SG $ denotes the Stop-Gradient operation. Therefore, in the second term, the loss is only used to update the codebook vectors. Due to the volume of the codebook space is dimensionless, the third term of Equation \ref{vq_loss} is called commitment loss, which constrains the range of values of encoder outputs $Z_e (x)$ by using penalty term $\beta$. \\

\begin{figure*}[htb]
	\begin{minipage}[b]{1.0\linewidth}
		\centering
		\centerline{\includegraphics[width=18cm]{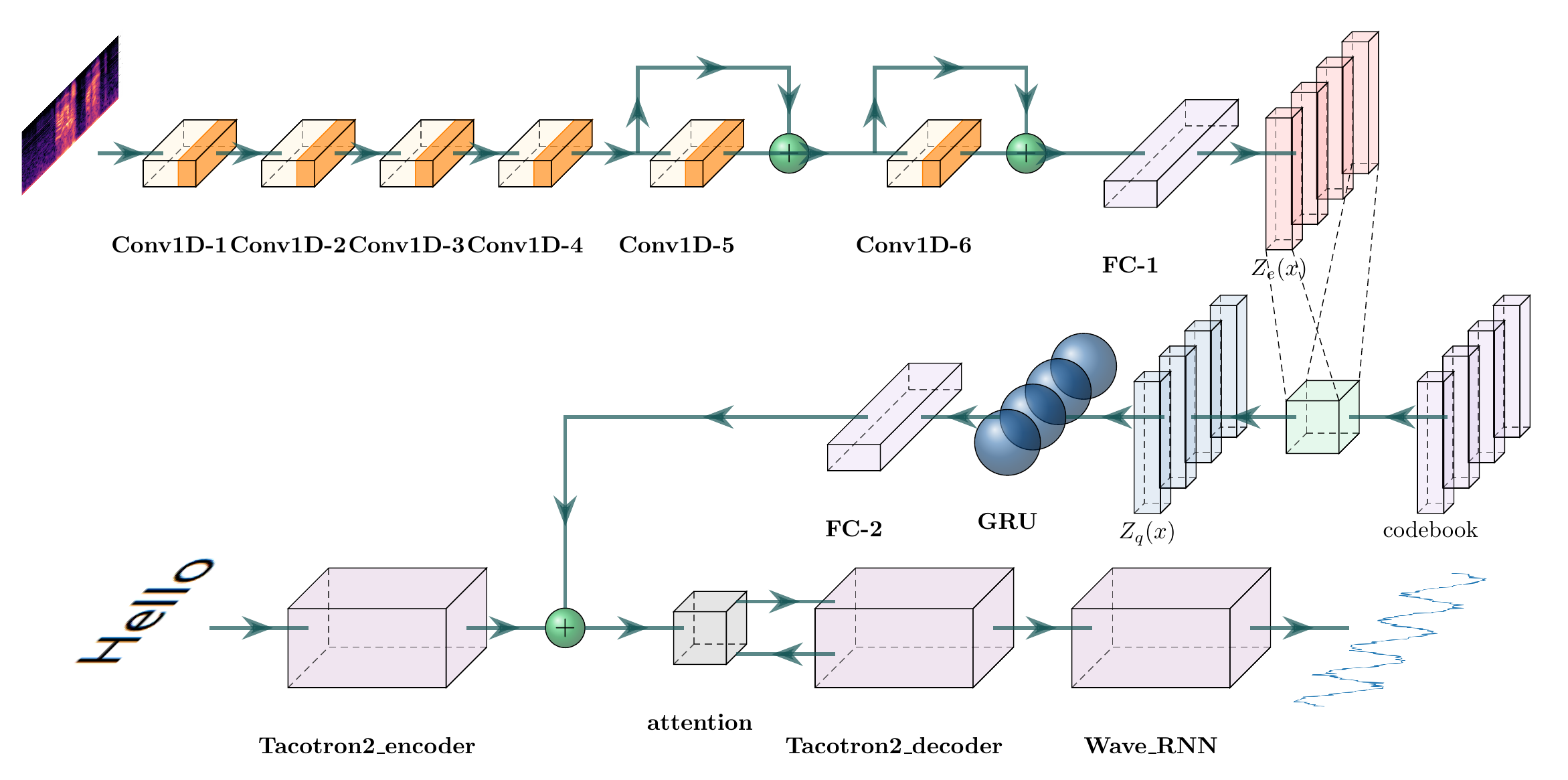}}
		\caption{VQ-encoder architecture. Each Conv1D layer is 512 units 1-D convolution, where 1st and 3rd Conv1D are kernel size 3 and stride 1 and the 2nd and 4th Conv1D are kernel size 4 and strides 2 for size reduction. ReLU activation is used after each convolution layer. The Conv1D layer in residual blocks is also 512 units 1-D convolution with kernel size 3 and stride 1. After the VQ process, the quantize latent vector $Z_q (x)$  with the size of 256$\times$16 is passed to a 128 unit GRU and a 256 unit fully connected layer to match the dimension of the text embedding.}
		\label{The_whole_architecture}
	\end{minipage}
\end{figure*}
\subsection{Whole architecture}
The whole architecture can be seen in Figure \ref{The_whole_architecture}, which consists of three different components: auxiliary encoder, Tacotron2 \cite{Shen2018}, and WaveRNN \cite{Kalchbrenner2018} vocoder. 
The Mel spectrum of the reference speech is passed to the auxiliary encoder. Then, the quantized vector is generated and passed to the Tacotron2 model to influences the synthesized speech. After that, the quantize vector is concatenated to the output of Tacotron2’s text encoder, which is passed to a Bahdanau attention modules \cite{Chorowski2015} and make the Tacotron2’s decoder generate Mel spectrum.  Finally, an WaveRNN \cite{Kalchbrenner2018} vocoder is used to reconstruct waveform from the Mel spectrum.
\begin{equation}
	\begin{aligned}
		L_{total}=&-E_{C\sim p_\phi (C|T),S\sim p_\theta (Q|R) } [\log p_\varphi (x | C ,Q)]\\
		&+L_{emb}+ L_{stop}   \label{total_loss}
	\end{aligned}
\end{equation}
The total loss of our model is shown in Equation \ref{total_loss}. In the first term of Equation \ref{total_loss}, $Q$ and $R$ denote quantize vector and reference audio respectively. The content vector $C$ given by input text $T$ is generated by the encoder of Tacotron2. So we can sample $Q$ and $C$ by the likelihood function $p_\theta (Q| R) $and $p_\phi (C| T)$, where $\theta$ and $\phi$ denote the parameters of text encoder and prosody encoder, respectively. After that, final output $X$ is sampled by $p_\varphi (x | C,S)$ which is a likelihood function of the decoder network. $L_{emb} $ is the second and third term of Equation \ref{vq_loss}, which aims to update the parameters of prosody encoder and latent codebook. $L_{stop}$ is the stop token loss which is to ensure the length of the Mel spectrum is correct.

\section{EXPERIMENTS and Discussion}
\label{sec:EXPERIMENT}
\subsection{Experimental setup}
The experiment includes two main content. First, to investigate the internal mechanism of VQ disentangle process, we use a numerical counter to accumulate the update values of latent codebook in different dimensions and find the relationship between the codebook and the real prosody. Second, to verify the effectiveness of VQ auxiliary encoder, we compare it with GST \cite{Wang2018} and VAE \cite{Akuzawa2018} auxiliary encoders. All the auxiliary encoders are concatenated to Tacotron2 \cite{Shen2018} and WaveRNN \cite{Kalchbrenner2018} to synthesis speech waveform.\\
We use two datasets to train and evaluate our models. First is Blizzard2013 \cite{Blizzard2013}, which includes 9733 utterances from stories for a total of 20 hours of audio, and second is the Chinese Standard Mandarin Speech Copus (CSMSC) \cite{CSMSC}, which contains 10000 utterances for a total of 12 hours of women's voices. All audio data are preprocessed as an 80-dimensional Mel spectrum, which is extracted with frame length 50ms and frameshift 12.5ms. 

\subsection{Why VQ can disentangle the prosody?}
\begin{figure}[htb]
	\begin{minipage}[b]{1.0\linewidth}
		\centering
		\centerline{\includegraphics[width=8.5cm, height=4cm]{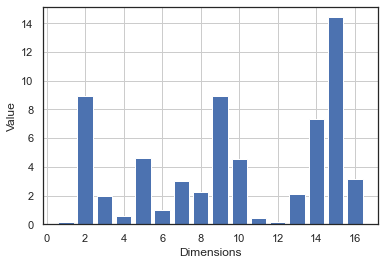}}
		\caption{The counter result of 16-dimension codebook. The dimensions with higher value have stronger correlation with prosody. }
		\label{vq_counter}
	\end{minipage}
\end{figure}

In order to find out why VQ latent variables can disentangle the reference speech and which dimensions of the VQ codebook affect the prosody of synthesized speech, we use a latent variable counter to observe the changes of codebook values. It records the absolute value of the average of the difference between the two latent variable matrices before and after each training step. After the training process, the accumulated value distribution is shown in the figure \ref{vq_counter}. 
The experiment shows that VQ process can learn different scales information and save it in corresponding codebook dimensions. In speech, the content is often the component with higher frequency, while the change of prosody is slower. Therefore, with the update of VQ codebook, they are disentangled and learned.\\
In this case, the top 3 dimensions are 2nd, 9th and 15th. By changing the values of these dimensions manually and listening the synthesized speech, we can recognize that they represent pitch, local pitch variation and speech velocity respectively. This phenomenon demonstrates we can build the relationship between the codebook and prosodies easily by observing the update rate of latent codebook. By contrast, other competition models lack of this ability, which would decrease the controllability.

\begin{figure}[htb]
	\begin{minipage}[b]{1.0\linewidth}
		\centering
		\centerline{\includegraphics[width=8.5cm]{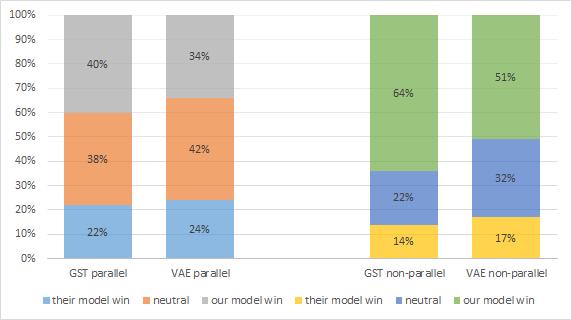}}
		\caption{Subjective evaluation results show our model have a higher recognizable identity. In the comparison to GST model, the number of people who feel our model better has an overwhelming advantage. In the comparison to VAE model, the evaluation opinions of VAE with good and neutral listening sense are increased compared with GST model, which shows that the listening sense of VAE is better than GST, but it is inferior to our model. }
		\label{ABX_results}
	\end{minipage}
\end{figure}

\subsection{Subjective evaluation}

For subjective evaluation, we take the ABX blind test \cite{Munson1950}, which compares two choices of sensory stimuli to identify detectable differences between them. Parallel transfer means the contents of the generated speech are the same as reference speech. Consequently, non-parallel denotes the contents are different. 50 audios and texts are selected from the test set for the style transfer task. The total number of judgers is 30. For each judger, they have three choices to evaluate the performance of generated audio: (1) A is better, (2) B is better, (3) neutral. The results are shown in Figure \ref{ABX_results}. From the diagram, our model outperforms the comparison models in both parallel and non-parallel transfer. 
\begin{table}[htb]
	\caption{ Objective evaluation results }  
	\label{Objective_evaluation}
	\begin{tabularx}{8cm}{llllX}  
		\hline                      
		Method & GPE & FFE & MCD & MOSNET  \\  
		\hline  
		GST & 0.47 & 0.45 & 15.0 & 3.15 \\
		VAE & 0.39 & 0.34 & 10.2 & 3.09 \\
		Our model & \textbf{0.18} & \textbf{0.16} &\textbf{ 8.7} & \textbf{3.19} \\
		\hline  
	\end{tabularx}  
\end{table} 

\subsection{Objective evaluation}

In objective evaluation, we also compare our model to the GST-encoder and VAE-encoder. At the beginning of the evaluation process, we use a pitch tracking algorithm \cite{Cheveigne?2002} to extract the pitch and voice decision from both of the reference audio and generated audio. Afterward, Gross Pitch Error (GPE) \cite{Tomohiro2008}, F0 Frame Error (FFE) \cite{Chu2009} and Mel-cepstral distortion (MCD) \cite{Kubichek1993} are employed to evaluate the generated speech, in which MCD is computed from the first 13 MFCCs. Beside them, we also use MOSNET \cite{Lo2019} to evaluate the quality of synthesized speech, which use a deep leaning network to simulate the human evaluations. The results are shown in Table \ref{Objective_evaluation}. The lower value of GPE, FFE, and MCD performs better and higher MOSNET is better. From the table, it is clear that our model outperforms the baseline models in all tests.
\begin{figure}[htb]
	\centering
	\subfloat[]{
		\includegraphics[width=4cm,height=2cm]{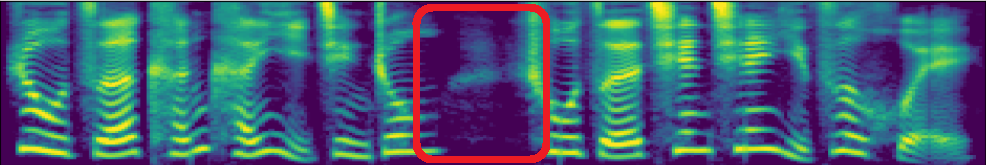}
	}
	\hfil
	\subfloat[]{
		\includegraphics[width=4cm,height=2cm]{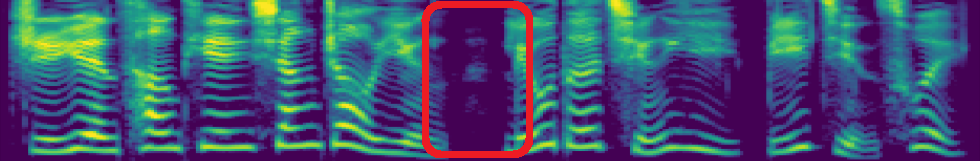}
	}
	\hfil
	\subfloat[]{
		\includegraphics[width=4cm,height=2cm]{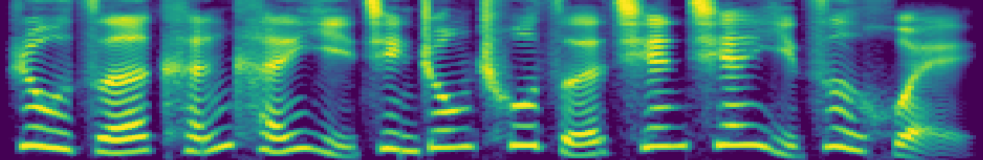}
	}
	\hfil
	\subfloat[]{
		\includegraphics[width=4cm,height=2cm]{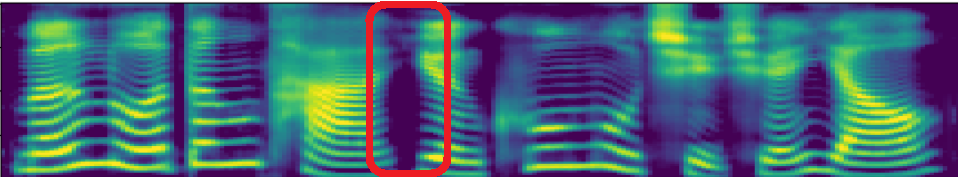}
	}
	
	\caption{Non-punctuation style transfer. The text used for generation is a Chinese poem with ten characters in a line. (a): the reference audio. (b): the synthesized audio of our proposed model. (c): the synthesized audio of the GST model, which can not learn the pause. (d): the synthesized audio of the VAE model, which pauses at a wrong position, where the target is after the 5th character but the generated pause position is before the 5th character. }
	\label{Non_punctuation_style_transfer}
\end{figure}

\begin{figure}[htb]
	\centering
	\subfloat[]{
		\includegraphics[width=4cm,height=2cm]{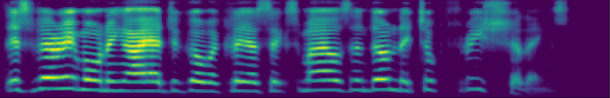}
	}
	\hfil
	\subfloat[]{
		\includegraphics[width=4cm,height=2cm]{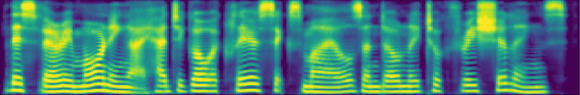}
	}
	\hfil
	\subfloat[]{
		\includegraphics[width=4cm,height=2cm]{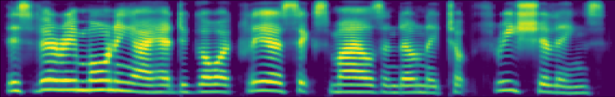}
	}
	\hfil
	\subfloat[]{
		\includegraphics[width=4cm,height=2cm]{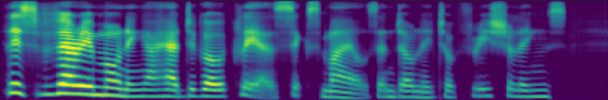}
	}
	\hfil
	\subfloat[]{
		\includegraphics[width=4cm,height=2cm]{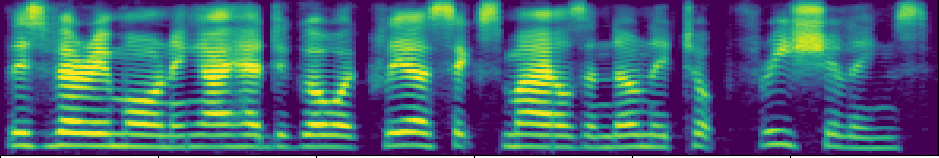}
	}
	\hfil
	\subfloat[]{
		\includegraphics[width=4cm,height=2cm]{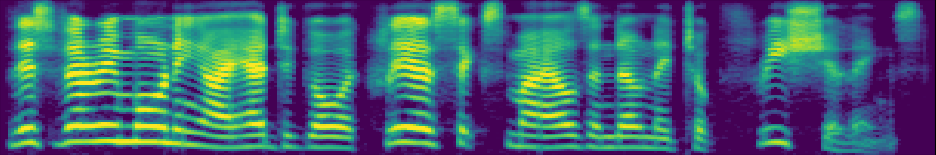}
	}	
	\caption{Style control by changing the values of the corresponding dimension of latent vectors. The speech is synthesized with the same text:” This very morning I had to go a clear six miles and only took three shillings.” (a) (b) exhibit the ability to control pitch by adjusting the 2nd dimension of latent vectors between -4 to 4. Spectrum diagram clearly shows pitch and corresponding harmonics changing from low to high. (c) and (d) demonstrate the speaking velocity varies by changing the value of the 15th dimension of the latent variable from -4 to 4. Obviously, (d) have higher velocity because it is shorter on time axis. (e) and (f) are the results of the 9th dimension value changing from -4 to 4. The change of local pitch is obvious.  This causes a variety in tone in the sense of listening. }
	\label{Style_control_by_changing_the_values}
\end{figure}

\subsection{Style transfer and control}
In addition, to verify the style transfer ability of our model, we design an interesting experiment named non-punctuation transfer task. Generally, the punctuation of the input text help the TTS model decide where should pause. Thus, if the input texts do not have punctuation, the existing models would fail to generate these pause segments. To demonstrate the power of our quantize reference encoder, we test the non-punctuation style transfer, which generates audio from text without punctuation, but the reference audio has corresponding pause segments. This experiment is a difficult task for the auxiliary encoder because it requires learning a subtle speaking style. The results are shown in Figure \ref{Non_punctuation_style_transfer}. Our proposed model can learn the break of speech between sentences. As a comparison, the synthesized audio of the GST and VAE model failed to learn the correct pause in the reference audio. 

Because of the advantages of disentangling speaking style, our model can control the style of generated speech by directly modifying the latent variables. In the style control experiment, we use 16-dimensional latent vectors to learn the style representation of the reference audio. The speaking velocity, pitch, and local pitch variation are controllable by adjusting relevant dimensions of the latent quantize vectors. The generated Mel spectrum are shown in Figure \ref{Style_control_by_changing_the_values}. By changing corresponding values, the synthesized speech is transformed to the wanted prosody. This function is particularly important in audio post editing, because it allows people to participate in audio synthesis and restore people's wishes more accurately.

\section{CONCLUSIONS}
\label{sec:CONCLUSIONS}

In this paper, we introduce a novel TTS model, which employs a quantize vector auxiliary encoder and can learn a disentangled prosody representation. The speech synthesized by our model performs better than current popular models in subjective listening and objective evaluations. Future work will focus on the study of the implicit mechanism of the disentangling process of vector quantize variables to improve the disentanglement and representation performance.

\bibliographystyle{IEEEbib}
\bibliography{refs}

\begin{thebibliography}{10}

\bibitem{Arik2017}
Sercan~{\"O}. Ar{\i}k, Mike Chrzanowski, Adam Coates, Gregory Diamos, Andrew
  Gibiansky, Yongguo Kang, Xian Li, John Miller, Andrew Ng, Jonathan Raiman,
  Shubho Sengupta, and Mohammad Shoeybi,
\newblock ``Deep voice: Real-time neural text-to-speech,''
\newblock in {\em Proceedings of the 34th International Conference on Machine
  Learning}, Doina Precup and Yee~Whye Teh, Eds. 06--11 Aug 2017, vol.~70 of
  {\em Proceedings of Machine Learning Research}, pp. 195--204, PMLR.

\bibitem{Wang2017}
Yuxuan Wang, R.J. Skerry-Ryan, Daisy Stanton, Yonghui Wu, Ron~J. Weiss, Navdeep
  Jaitly, Zongheng Yang, Ying Xiao, Zhifeng Chen, Samy Bengio, Quoc Le, Yannis
  Agiomyrgiannakis, Rob Clark, and Rif~A. Saurous,
\newblock ``{Tacotron: Towards End-to-End Speech Synthesis},''
\newblock in {\em Proc. Interspeech 2017}, 2017, pp. 4006--4010.

\bibitem{Shen2018}
J.~Shen, R~Pang, RJ~Weiss, M.~Schuster, and Y.~Wu,
\newblock ``Natural tts synthesis by conditioning wavenet on mel spectrogram
  predictions,''
\newblock in {\em ICASSP 2018 - 2018 IEEE International Conference on
  Acoustics, Speech and Signal Processing (ICASSP)}, 2018.

\bibitem{Ren2019}
Yi~Ren, Yangjun Ruan, Xu~Tan, Tao Qin, Sheng Zhao, Zhou Zhao, and Tie-Yan Liu,
\newblock ``Fastspeech: Fast, robust and controllable text to speech,''
\newblock in {\em Advances in Neural Information Processing Systems},
  H.~Wallach, H.~Larochelle, A.~Beygelzimer, F.~d\textquotesingle
  Alch\'{e}-Buc, E.~Fox, and R.~Garnett, Eds. 2019, vol.~32, Curran Associates,
  Inc.

\bibitem{Ren2021}
Yi~Ren, Chenxu Hu, Xu~Tan, Tao Qin, Sheng Zhao, Zhou Zhao, and Tie-Yan Liu,
\newblock ``Fastspeech 2: Fast and high-quality end-to-end text to speech,''
\newblock in {\em International Conference on Learning Representations}, 2021.

\bibitem{Wang2018}
Yuxuan Wang, Daisy Stanton, Yu~Zhang, RJ-Skerry Ryan, Eric Battenberg, Joel
  Shor, Ying Xiao, Ye~Jia, Fei Ren, and Rif~A. Saurous,
\newblock ``Style tokens: Unsupervised style modeling, control and transfer in
  end-to-end speech synthesis,''
\newblock in {\em Proceedings of the 35th International Conference on Machine
  Learning}, Jennifer Dy and Andreas Krause, Eds. 10--15 Jul 2018, vol.~80 of
  {\em Proceedings of Machine Learning Research}, pp. 5180--5189, PMLR.

\bibitem{Kwon2019}
Ohsung Kwon, Inseon Jang, ChungHyun Ahn, and Hong-Goo Kang,
\newblock ``An effective style token weight control technique for end-to-end
  emotional speech synthesis,''
\newblock {\em IEEE Signal Processing Letters}, vol. 26, no. 9, pp. 1383--1387,
  2019.

\bibitem{Wu2020}
P.~Wu, Z.~Ling, L.~Liu, Y.~Jiang, H.~Wu, and L.~Dai,
\newblock ``End-to-end emotional speech synthesis using style tokens and
  semi-supervised training,''
\newblock in {\em 2019 Asia-Pacific Signal and Information Processing
  Association Annual Summit and Conference (APSIPA ASC)}, 2020.

\bibitem{Akuzawa2018}
K.~Akuzawa, Y.~Iwasawa, and Y.~Matsuo,
\newblock ``Expressive speech synthesis via modeling expressions with
  variational autoencoder,''
\newblock in {\em Interspeech}, 2018.

\bibitem{Sun2020}
G.~Sun, Y.~Zhang, R.~J. Weiss, Y.~Cao, H.~Zen, and Y.~Wu,
\newblock ``Fully-hierarchical fine-grained prosody modeling for interpretable
  speech synthesis,''
\newblock in {\em ICASSP 2020 - 2020 IEEE International Conference on
  Acoustics, Speech and Signal Processing (ICASSP)}, 2020.

\bibitem{Zhang2019}
Y.~J. Zhang, S.~Pan, L.~He, and Z.~H. Ling,
\newblock ``Learning latent representations for style control and transfer in
  end-to-end speech synthesis,''
\newblock in {\em ICASSP 2019 - 2019 IEEE International Conference on
  Acoustics, Speech and Signal Processing (ICASSP)}, 2019.

\bibitem{An2019}
Xiaochun An, Yuxuan Wang, Shan Yang, Zejun Ma, and Lei Xie,
\newblock ``Learning hierarchical representations for expressive speaking style
  in end-to-end speech synthesis,''
\newblock in {\em 2019 IEEE Automatic Speech Recognition and Understanding
  Workshop (ASRU)}, 2019, pp. 184--191.

\bibitem{Hsu2019}
Wei-Ning Hsu, Yu~Zhang, Ron Weiss, Heiga Zen, Yonghui Wu, Yuan Cao, and Yuxuan
  Wang,
\newblock ``Hierarchical generative modeling for controllable speech
  synthesis,''
\newblock in {\em International Conference on Learning Representations}, 2019.

\bibitem{Oord2017}
Aaron van~den Oord, Oriol Vinyals, and koray kavukcuoglu,
\newblock ``Neural discrete representation learning,''
\newblock in {\em Advances in Neural Information Processing Systems}, I.~Guyon,
  U.~V. Luxburg, S.~Bengio, H.~Wallach, R.~Fergus, S.~Vishwanathan, and
  R.~Garnett, Eds. 2017, vol.~30, Curran Associates, Inc.

\bibitem{Razavi2019}
Ali Razavi, Aaron van~den Oord, and Oriol Vinyals,
\newblock ``Generating diverse high-fidelity images with vq-vae-2,''
\newblock in {\em Advances in Neural Information Processing Systems},
  H.~Wallach, H.~Larochelle, A.~Beygelzimer, F.~d\textquotesingle
  Alch\'{e}-Buc, E.~Fox, and R.~Garnett, Eds. 2019, vol.~32, Curran Associates,
  Inc.

\bibitem{Wu2020a}
Da-Yi Wu, Yen-Hao Chen, and Hung yi~Lee,
\newblock ``{VQVC+: One-Shot Voice Conversion by Vector Quantization and U-Net
  Architecture},''
\newblock in {\em Proc. Interspeech 2020}, 2020, pp. 4691--4695.

\bibitem{Kalchbrenner2018}
Nal Kalchbrenner, Erich Elsen, Karen Simonyan, Seb Noury, Norman Casagrande,
  Edward Lockhart, Florian Stimberg, Aaron van~den Oord, Sander Dieleman, and
  Koray Kavukcuoglu,
\newblock ``Efficient neural audio synthesis,''
\newblock in {\em Proceedings of the 35th International Conference on Machine
  Learning}, Jennifer Dy and Andreas Krause, Eds. 10--15 Jul 2018, vol.~80 of
  {\em Proceedings of Machine Learning Research}, pp. 2410--2419, PMLR.

\bibitem{Chorowski2015}
Jan~K Chorowski, Dzmitry Bahdanau, Dmitriy Serdyuk, Kyunghyun Cho, and Yoshua
  Bengio,
\newblock ``Attention-based models for speech recognition,''
\newblock in {\em Advances in Neural Information Processing Systems},
  C.~Cortes, N.~Lawrence, D.~Lee, M.~Sugiyama, and R.~Garnett, Eds. 2015,
  vol.~28, Curran Associates, Inc.

\bibitem{Blizzard2013}
``https://www.synsig.org/index.php/blizzard\_challenge\_2013,'' .

\bibitem{CSMSC}
``https://www.data-baker.com/open\_source.html,'' .

\bibitem{Munson1950}
W.~A. Munson,
\newblock ``Standardizing auditory tests,''
\newblock {\em The Journal of the Acoustical Society of America}, vol. 22, no.
  5, pp. 675--675, 1950.

\bibitem{Cheveigne?2002}
A~De Cheveigne? and H.~Kawahara,
\newblock ``Yin, a fundamental frequency estimator for speech and music,''
\newblock {\em Journal of the Acoustical Society of America}, vol. 111, no. 4,
  pp. 1917--30, 2002.

\bibitem{Tomohiro2008}
Tomohiro, Nakatani, , , Shigeaki, Amano, , , Toshio, Irino, , , and Kentaro
  and,
\newblock ``A method for fundamental frequency estimation and voicing decision:
  Application to infant utterances recorded in real acoustical environments,''
\newblock {\em Speech Communication}, vol. 50, no. 3, pp. 203--214, 2008.

\bibitem{Chu2009}
W.~Chu and A.~Alwan,
\newblock ``Reducing f0 frame error of f0 tracking algorithms under noisy
  conditions with an unvoiced/voiced classification frontend,''
\newblock in {\em IEEE International Conference on Acoustics}, 2009.

\bibitem{Kubichek1993}
R.~Kubichek,
\newblock ``Mel-cepstral distance measure for objective speech quality
  assessment,''
\newblock in {\em Communications, Computers and Signal Processing, 1993., IEEE
  Pacific Rim Conference on}, 1993.

\bibitem{Lo2019}
CC~Lo, S.~W. Fu, W.~C. Huang, X.~Wang, and H.~M. Wang,
\newblock ``Mosnet: Deep learning based objective assessment for voice
  conversion,''
\newblock in {\em Interspeech 2019}, 2019.

\end{thebibliography}

\end{document}